# Performance of the CREAM calorimeter in accelerator beam test

Y. S. Yoon[a], H. S. Ahn[b], , M. G. Bagliesi[c], , G. Bigongiari[c], O. Ganel[b], J. H. Han[d], H. J. Hyun[d], J. A. Jeon[d], T. G. Kang[b], H. J. Kim[e], K. C. Kim[b], J. K. Lee[d], M. H. Lee[b], L. Lutz[b], P. Maestro[c], A. Malinine[b], P. S. Marrocchesi[c], S. W. Nam[d], H. Park[e], I. H. Park[d], N. H. Park[d], E. S. Seo[a,b], R. Sina[b], J. Wu[b], J. Yang[d], R. Zei[c] and S. Y. Zinn[b]
*(a) Dept. of Physics, University of Maryland, College Park, MD 20742, USA*
*(b) Inst. For Phys. Sci. and Tech., University of Maryland, College Park, MD 20742, USA*
*(c) Dept. of Physics, University of Siena and INFN, Via Roma 56, 53100 Siena, Italy*
*(d) Dept. of Physics, Ewha Womans University, Seoul, 120-750, Republic of Korea*
*(e) Dept. of Physics, Kyungpook National University, Taegu, 702-701, Republic of Korea*
Presenter: Young Soo Yoon (ysy@umd.edu), usa-yoon-Y-abs3-he15-poster

The CREAM calorimeter, designed to measure the spectra of cosmic-ray nuclei from under 1 TeV to 1000 TeV, is a 20 radiation length ($X_0$) deep sampling calorimeter. The calorimeter is comprised of 20 layers of tungsten interleaved with 20 layers of scintillating fiber ribbons, and is preceded by a pair of graphite interaction targets providing about 0.42 proton interaction lengths ($\lambda_{int}$). The calorimeter was placed in one of CERN's SPS accelerator beams for calibration and testing. Beams of 150 GeV electrons were used for calibration, and a variety of electron, proton, and nuclear fragment beams were used to test the simulation model of the detector. In this paper we discuss the performance of the calorimeter in the electron beam and compare electron beam data with simulation results.

## 1. Introduction

The cosmic-ray spectrum follows a decreasing power law, with the incident flux dropping by roughly a factor of 50 for every ten-fold increase in the threshold energy. Thus, of all particles above 1 TeV, only 2% are above 10 TeV, and only 0.04% above 100 TeV. It is for this reason that the limitation on the energy reach of flight experiments measuring these particles is mostly statistical. The CREAM instrument was designed to directly measure the charge and energy of cosmic-ray nuclei, from H to Fe, for energies ranging from under 1 TeV to 1000 TeV [1]. CREAM incorporates both a calorimeter and a transition radiation detector (TRD) to measure particle energies (TRD for Z>3), and a timing-based charge detector (TCD) and a silicon charge detector (SCD) to measure particle charge or ID. The independent measurements allow verification and cross-calibration for large sub-samples of events collected. The calorimeter uses a sampling scheme with tungsten absorber to minimize the thickness required for a given depth in $X_0$, allowing a greater collection power, or geometry factor (GF). The graphite targets, with their relatively low Z number (6 vs. 74 for tungsten), increase the fraction of nuclei interacting early enough to allow a reasonable reconstruction of their incident energy, while minimizing the weight, and having a minimal impact on the 'age' of the shower development regardless of the depth of graphite the shower secondaries must traverse before reaching the active scintillating fiber layers. The CREAM payload was flown for about 42 days suspended under a Long Duration Balloon (LDB) launched from McMurdo Station, Antarctica in December 2004.

## 2. CREAM calorimeter module

Figure 1 shows a schematic cross sectional view of various components in the CREAM calorimeter module. The calorimeter is made of 20 layers of one radiation length ($X_0$) thick tungsten plates and fifty 1 cm wide fiber ribbons each made of nineteen 0.5 mm diameter BCF-12 scintillating fibers covering 0.25 m$^2$. The scintillation light from the fiber ribbons is transmitted via a light mixer and a bundle of clear fibers to a set of 73-pixel hybrid photo diodes (HPDs). The HPDs combine a high channel count (73 each), low weight (~30 g), low power consumption (~0.8W each, including HV supply and front-end electronics), and have both high uniformity between pixels (RMS of ~5%), and a linear dynamic range of about 1:10$^6$. To cover both the low energy signals on the periphery of showers, and the high density energy signal at the core of the highest energy showers, a dynamic range of 1:200,000 was required. Since the front end electronics has a limited dynamic range (~1:1000 per channel), a means was needed to split the signal into several ranges, with some overlap for calibration purposes. The clear fibers carrying the signal from each scintillating fiber ribbon were thus divided into three different sub-bundles, with 37 thin fibers for the low energy range, 5 for the middle range, and 1 for the high range. These sub-bundles were glued into plastic cookies to position them against separate HPD pixels. In addition to the optical signal splitting, the signals from the middle and high ranges were attenuated using neutral density filters glued on the surface of the cookies, with transmission efficiencies of ~50% and ~16%, respectively, completing the dynamic range needed.

## 3. Calibration method

With a total of 1000 fiber ribbons in 20 layers, each with three readout ranges, calibration is of critical importance in correctly reconstructing shower energy. There are two steps to this calibration. First, one must equalize the low ranges of the different ribbons. Then, one must inter-calibrate the mid- to low-range and the high- to mid-range. The equalization process required scanning the surface of the calorimeter with beams of 150 GeV electrons, and for each ribbon, comparing the measured signal in ADC counts for those events where that ribbon was in the middle of the beam profile, with the energy deposit from simulated 150 GeV electron events with the same incident positions. As seen in Figure 1, the mean measured signal for a typical ribbon was 59.86 ADC units, while the simulated energy deposit for that ribbon had a fit mean of 49.59 MeV. This gives rise to a calibration constant of 0.8284 MeV/ADC count. The CREAM calorimeter is designed for multi-TeV showers, and thus has an electron sampling fraction of ~0.5%. This results in an energy deposit too low to be of use in calibrating the ribbons of the first and last few layers when using 150 GeV electrons. To address the calibration of the first layers, 2.5 cm thick (~5 $X_0$) lead bricks were placed in front of the graphite targets for a second scan to move the shower maximum closer to the top of the calorimeter. To calibrate the last few layers, the detector was rotated such that the beam was incident through its bottom, and the same lead bricks were placed before the aluminum honeycomb pallet holding the stack. To compensate for the missing 2$X_0$ of the targets and the tungsten layer above the first scintillator layer, two additional tungsten plates (1 $X_0$ thick each) were placed between the lead bricks and the pallet. A third scan in this configuration moved the shower maximum to the bottom portion of the calorimeter and allowed the lower layer ribbons to be calibrated. Figure 2(a) shows the distribution of the 1000 equalization

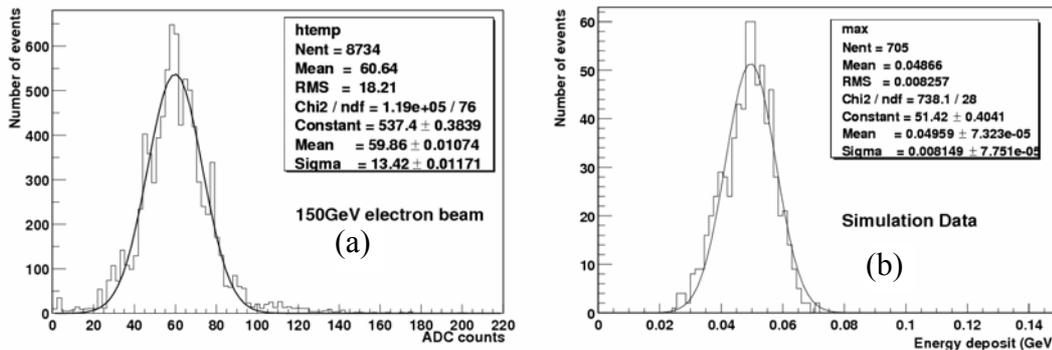

**Figure 1**. (a) A distribution of ADC counts for the maximum ribbon of a layer in 150 GeV electron beam data; (b) A distribution of energy deposit in the same ribbon in simulation data

constants obtained through the above process. The simulation included such effects as photon statistics, incoherent noise, and coherent noise. After applying the calibration constants for the 1000 low range readout channels, the beam data showed very good agreement with the simulation results (see Figure 2(b)).

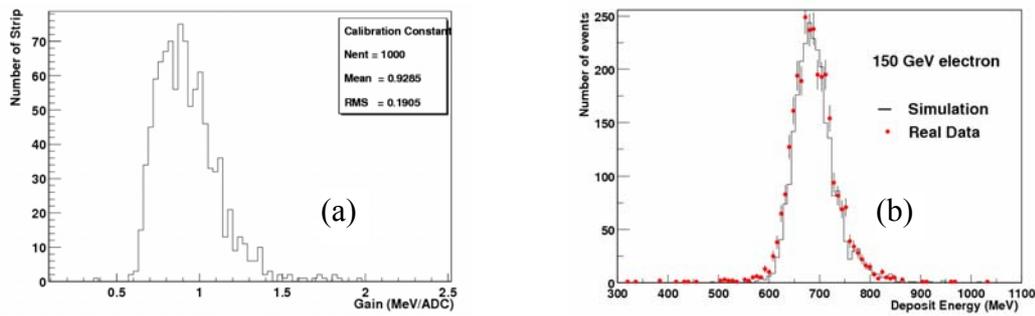

**Figure 2.** (a) Distribution of calibration constants used in this analysis; (b) Calorimeter energy distributions for 150 GeV electrons from simulation and beam data.

For reasons of package volume, weight, and power consumption, the CREAM calorimeter uses a custom, ASIC-based readout system, with 32 channels per ASIC. This system has low noise, but some of this noise is coherent. In addition, with the calorimeter effective Moliere radius at only ~1.12 cm, it is not useful to sum over the full 50 cm width of the calorimeter. To optimize the signal to noise ratio, the energy resolution was assessed vs. the number of ribbons summed over per layer, with the best resolution obtained for 5 ribbons per layer (total of 100 ribbons). With this method, the calorimeter reconstructed energy and the resolution were plotted vs. the electron beam energy (Figure 3). The reconstructed energy proved to be linear with the beam energy with a slope of 5.036 MeV/GeV or 0.5% sampling. The offset from the origin is such that the reconstructed signal would extrapolate to zero when the incident energy is 3 GeV, a very small effect for a detector designed to measure multi-TeV showers, and likely related to the fact that very small signals are discarded to minimize the telemetry bandwidth needed to transmit the data in flight from the payload to the ground in near real time. The energy resolution of ~6% at 150 GeV and higher is more than sufficient for calibration purposes, where a large number of events is used for each calibration constant. Since the experiment is intended to measure shower energies of very high energy nuclei, calibration accuracy is the only significance of the electron energy resolution.

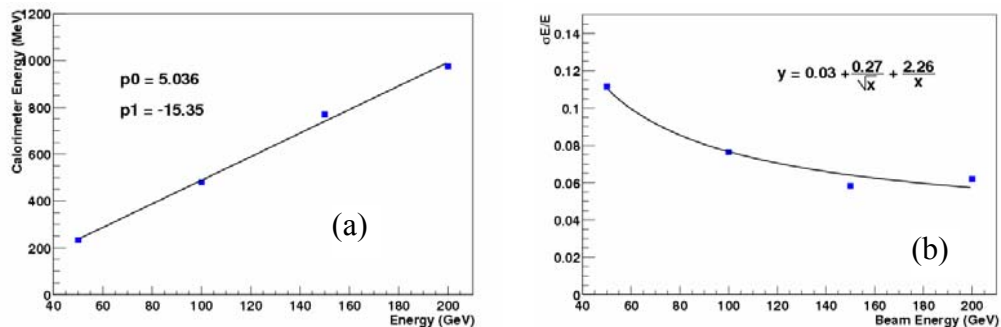

**Figure 3.** (a) Calorimeter energy vs. electron beam energy (b) energy resolution

## 2. Position and angle uniformity

Utilizing a fine scan, with a pitch of 0.2 mm, the reconstructed energy measurement and resolution were assessed as a function of beam position. As shown in Figure 4, the measured energy is independent of incidence position within ±3%, and the resolution is independent of incidence position within ±6%. Further, the variations between different positions do not coincide with ribbon boundaries, and are thus most likely statistical in nature. Figure 5 shows the results of similar measurements for an incident angle scan, demonstrating the detector is not sensitive to incident angles for 150 GeV electrons.

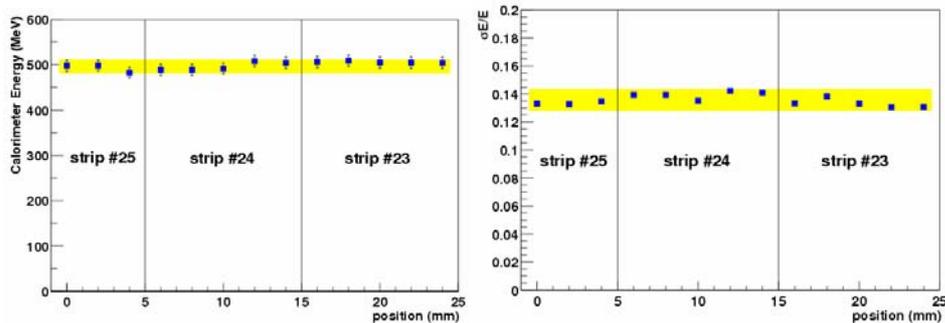

**Figure 4.** (a) Calorimeter energy and (b) energy resolution with respect to the electron beam injection points.

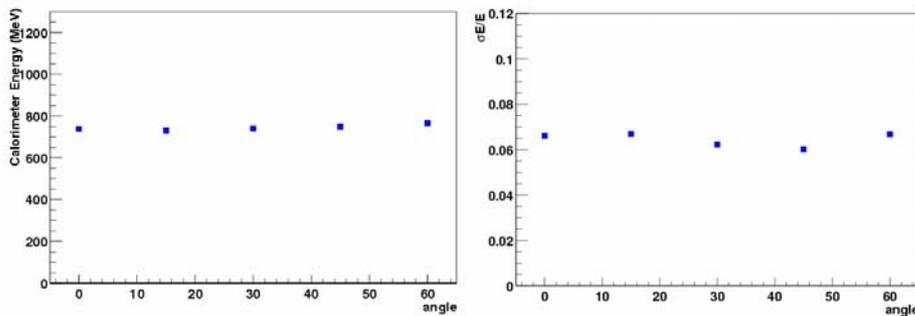

**Figure 5.** (a) Calorimeter energy and (b) energy resolution with different electron beam incident angles.

## 3. Conclusions

The CREAM calorimeter was designed to measure the spectra of high energy cosmic-ray nuclei. The calorimeter was placed in a CERN electron beam for calibration and to test its performance. The results show good agreement with simulation, and prove that the detector is linear for electrons, has an electron resolution sufficient for calibration, and is uniform down to a 2 mm pitch. The measured energy and resolution were not sensitive to the incident angle for 150 GeV electrons.

## 4. Acknowledgements

This work was supported by NASA. The authors thank NASA/WFF, National Scientific Balloon Facility, NSF Office of Polar Programs, and Raytheon Polar Service Company for the successful balloon launch, flight operations, and payload recovery. The authors also thank the support of CERN for the excellent beam test facilities and operations.